\documentclass[twocolumn,english,aps,prl,groupedaddress,superscriptaddress]{revtex4}
\usepackage{graphicx,epsfig,units}
\usepackage{xcolor} 
\usepackage{soul} 
\usepackage{amsmath,amsfonts,mathrsfs,amsbsy,bm,babel}

\begin{document}

\title{Roughness as a measurement of random disorder}

\author{M.~Ramazanoglu}

\affiliation{Physics Engineering Department, Istanbul Technical University, 34469, Maslak, Istanbul, Turkey}

\author{R. ~Salci}
\affiliation{Physics Engineering Department, Istanbul Technical University, 34469, Maslak, Istanbul, Turkey}

\begin{abstract}
We have studied the quenched random disorder (QRD) effects created by aerosil dispersion in octylcyanobiphenyl (8CB) liquid crystal (LC)
using Atomic Force Microscopy (AFM) technique. Gelation process in the 8CB+aerosil gels yields a QRD network which also changes the surface topography.
By increasing the aerosil concentration, the original smooth pattern of LC sample surfaces is suppressed by creating a fractal aerosil surface effect;
these surfaces become more porous, rougher with more and bigger crevices. The dispersed aerosil also 
serves as pinning centers for the liquid crystal molecules.
It has been observed that via the diffusion-limited-aggregation process, aerosil nano-particles yield a fractal like surface pattern for the less disordered samples.
As the aerosil dispersion increases, the surface can be described by more aggregated regions which also introduce more roughness. Using this fact, 
we have shown that there is a net correlation between the short-ranged x-ray peak widths (-the results of previous x-ray diffraction experiments-) 
and the calculated surface roughness. In other words, we have shown that these QRD gels can also be characterized by their surface roughness values.

\end{abstract}

\date{March 19,2017}
\maketitle

\section{Introduction}

Since liquid crystals (LCs) show rich phase transition capabilities within several mesogenic phases, they have managed to be a trendy subject for science
where disorder effects are studied within controlled random forces. For almost two decades, aerogel and aerosil nano-particles were used to create 
the random disorders
in LC phases \cite{Iannacchione,Haga}. In these previous works, the aim was to study the randomness imposed on the LC order parameters where the amount of dispersed silica nano-particles simply served as the
control mechanism of the created QRD. 

The first experiments were conducted on LC+aerogel samples using mainly heat capacity and x-ray scattering techniques \cite{Germano, Sungil}. It was observed from
these measurements that the aerogel dispersion couples with the nematic director and smectic order   
creating a random gel network. Later, the quenched disorder effects of aerosil were found to be more interesting than the ones created by aerogels, such that
many subsequent LC+aerosil experiments followed \cite{Iannacchione, Haga}. The LC+aerosil systems are particularly important because very low random disorder effects can be 
created within these gels. 
This is achieved with the hydroxyl groups surrounding the silica nano-particles. Each hydroxyl covered silica sphere surfaces affects one an other 
creating fractal links of
solid nano-particles in  the LC host environment. These links yield a weak network giving flexibility to the LC+aerosil gels. 

In these systems the coupling between nematic, smectic order parameters (amplitude and the one dimensional density phase)
and the aerosil nano-particles can be used 
to describe the quenched randomness seen in Type II superconductors and other magnetic systems where one single order parameter cannot be enough
to model the physics. As a result a more complex anisotropic critical behavior is studied in LC+aerosil gels.

Here we aim to study the surface characteristics of LC+aerosil samples using Atomic Force Microscopy (AFM) techniques. The study of the surface 
topography has the advantage of correlating the results to the well-known characteristics of these gels. Randomly pinned LC orders, so far, have been
investigated
by several characteristic parameters including $\xi_{RD}$; random disorder correlation length, $\delta$H; enthalpy and $\Delta$C$_p$; heat capacity.
These parameters are measured by x-ray diffraction and calorimetry spectroscopy techniques, respectively.
Different from these, we will use the surface roughness to characterize the same effect for the first time.

To reach these goals, 8CB (4-cyano-4-octylbiphenyl) LC was selected. This way, a second 
order phase transition in which short ranged correlations in smectic A phase created by the influence of aerosil dispersion will be studied in this article.
All of the measurements shown and discussed here focus on the topographical changes on  the sample surfaces. In that sense, this can be categorized as a surface
science study of LC+aerosil gels.
However the central idea in this article is the study of the correlation between the known
short-ranged smectic order characteristics and the surface topography of the gel samples.

\section{Experiment Details}
The 8CB LC was purchased from Frinton Lab \cite{Frinton} and we performed our sample preparation steps without any purification process. 
Hydrophilic aerosil was obtained 
from Evonik Corp.\cite{Degussa}. In this article only type 300 which is $\sim$ 7 nm in diameter silica nano-particles were used.The Brunauer-Emmet-Teller (BET) surface 
area for these nano-particles is listed as 300 m$^2$g$^-1$\cite{Degussa}. Since they are very susceptible to water contamination, a drying process at T$\simeq$500 K
under ~10$^-2$ atm vacuum was used for several days prior to mixing with LCs. In this study the samples were distinguished with a parameter
known as aerosil mass density $\rho_s$ which can be derived from 

\begin{equation}
 \frac{1}{\rho_s}=\frac{1}{\rho}-\frac{1}{\rho_{aerosil}}                 \label{rhos}
\end{equation}
where 
\begin{equation}
 \rho=\frac{m_{aerosil}}{V_{tot}}                                        \label{rho}
\end{equation} and $\rho_{aerosil}$=2.2 g/cm$^3$ \cite{Degussa}. 
Here; V$_{tot}$=V$_{LC}$+V$_{aerosil}$.
Thus, Eq.\ref{rhos} can also be simplified as;
\begin{equation}
 \rho_s=\frac{m_{aerosil}}{V_{LC}}                                       \label{rhosII}
\end{equation} 

Eq. \ref{rhosII} is more useful for categorizing the sample disorder strength which is simply the ratio between mass of the aerosil and the volume of the LC. 
Stoichiometric amounts of 8CB and the 
aerosil were mixed in high purity ethanol and sonicated at $\sim$ 300 K for around half an hour. This way high quality 8CB+aerosil mixtures were 
successfully obtained \cite{mmt,mmt1,mmt2,Freelon}. The mixtures were then placed on a hot plate which was held at 310 $\pm$0.5 K constant temperature.
The rest of the drying, or in other
words gelation, process was performed at a temperature $\sim$ 307 K  close to the isotropic phase of the LC. 
For AFM measurements samples were 
transferred on microscope glass specimen holders. Drying of the high aerosil concentration samples produced small cracks,  where it was possible to hold them.
These samples were placed on glass holders using tweezers. However, low aerosil concentration samples; $\rho_s$=0.1 and 0.2  were collected with a spatula
and an adequate amount of sample
was placed on the glass -or a Si wafer surface-. The gelation procedure creates mainly two groups of LC+aerosil gels depending on the aerosil amount \cite{Iannacchione}. The samples with $\rho_s\succ\sim~$0.2 g/cm$^3$
are found to be more rigid and their physical attributes resemble the LC+aerogel samples \cite{Haga}. 
During the  process of sample transfer, samples were carefully held at the same drying temperature lest the LC+aerosil network  
be mechanically perturbed and put into temperature stress. 
This way, we also avoided any accidental crystallization and phase separation issues in the 
samples. In that sense,  8CB LC is one of the best and easiest LC samples to work with because it crystallizes at T$_C\sim$290 K ($\sim$ 17 $^o$C); 
lower than common room temperatures.
It has nematic to smectic $A$  and isotropic to nematic phase transitions at T$_{NA}\sim$ 306.97 K and T$_{NI}\sim$ 313.98 K , respectively \cite{Leheny,Sungil}.  
The AFM scans were conducted at the ITUNANO Laboratory clean room facility where the temperature and 
humidity were held at 296.15 K and $\sim$35 \% constantly \cite{AFM}. This temperature is just 6 K over the crystallization temperature of the 8CB 
so that the AFM scans were conducted on the well-formed and short-range ordered smectic $A$ phases of the 8CB+aerosil gels surfaces. The surface
topography was investigated by a profoliometer \cite{profoliometer} for the two highest aerosil mass density samples. The AFM scans were analyzed by 
NMI Image Analyzer ver 1.4 software\cite{AFM}.

\section{Results}
\begin{figure}
\includegraphics[width=1\linewidth]{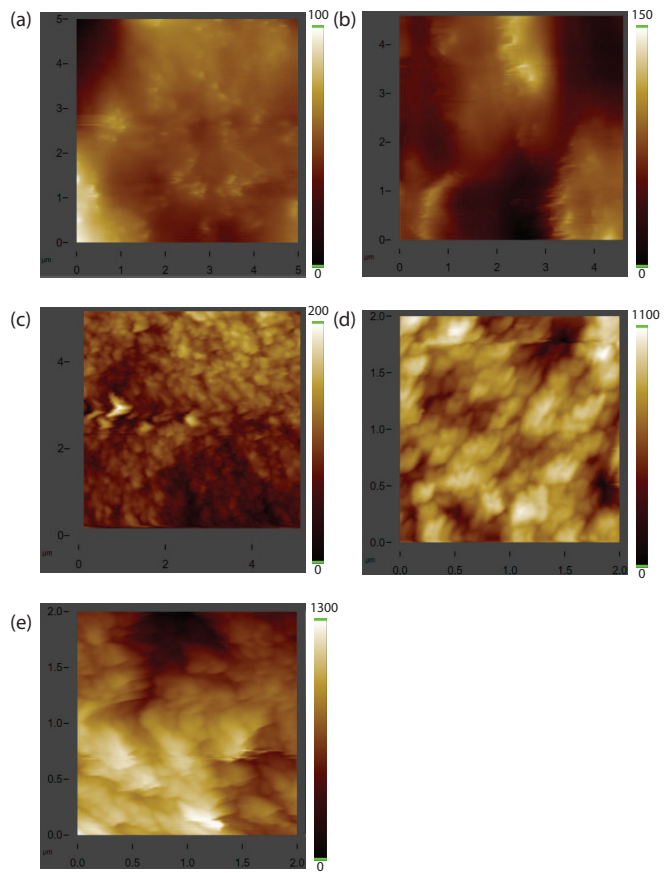}
\caption{\footnotesize AFM scans of sample surfaces ranging from $\rho_s$=0.104 (panel (a)) to $\rho_s$=0.647 (panel (e)) g/cm$^3$. These are 256x256 pixel 
resolution 5x5 $\mu$m$^2$ for panel (a to c) and 2x2 $\mu$m$^2$ for panel (d and e) in size tapping mode data. Between panel (a) to panel (e), there is 
a net change both in the smoothness and the fractal structure of the surface as discussed in the text.}
\label{fig1}
\end{figure}

The atomic force microscopy (AFM) data are shown in Fig.\ref{fig1} for 5 different pictures of 5 different aerosil mass density ($\rho_s$) values. 

\begin{figure}
\includegraphics[width=1\linewidth]{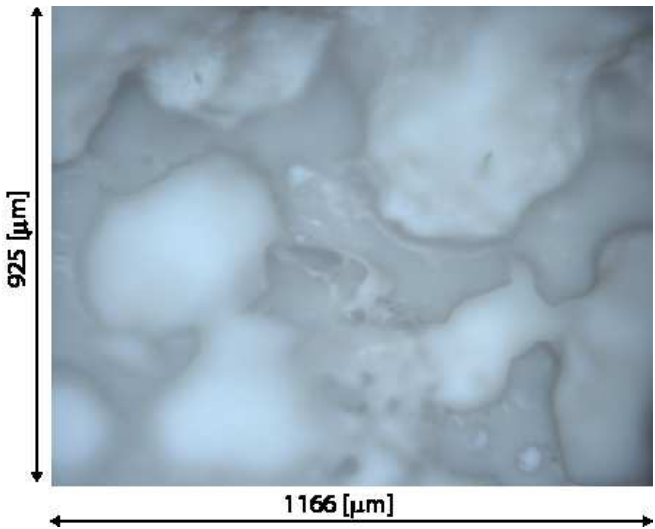}
\caption{\footnotesize Profilometer microscope picture \cite{profoliometer}. This is the actual surface image of the highest disordered strength sample;
$\rho_s$=0.647 g/cm$^3$.  }
\label{fig2}
\end{figure}

Here the data was only interpreted with the surface topography value; surface roughness $<R>$ which can be given as 
\begin{equation}
 <R>=\frac{1}{MN} \sum_{h=0}^{M-1}\sum_{k=0}^{M-1}\mid z(h,k)\mid                \label{R}
\end{equation}\cite{AFM}. 
In Eq.\ref{R}, $z(h,k)$ denotes the normalized height distribution data where the mean surface height distribution is subtracted \cite{Raposo}. 
Both non-contact and tapping modes were used however the pictures shown in Fig.\ref{fig1} were obtained with the tapping mode only.  
Surface roughness values of the samples were studied with respect to random disorder strength; $\rho_s$ which is given by Eq.\ref{rhosII}. 

As seen from $\rho_s$=0.105 g/cm$^3$ ; panel (a) of Fig.\ref{fig1}, the surface is mostly composed of LC where a smooth surface topography  
is observed. The small white colored sharp fractal like regions were created by aerosil dispersion in the LC. The dispersion of aerosil was 
thought to create
a fractal network in the LC medium. This was previously discussed extensively \cite{Germano}. These sharp white small areas were the realization of 7 nm 
diameter aerosil spheres which come together to form diffused-aggregated regions $\sim$50 nm in size for $\rho_s$=0.1 g/cm$^2$ sample shown
in panel (a). This corresponds to $\sim$ 7 silica particles.
The hydrogen coated aerosil used in this experiment would form this kind of chemically
fused fractal networks which are weakly attached to each other \cite{Germano}. 
Therefore void length scales are formed for the LC molecules which in turn disorder their
nematic features creating a random disorder effect on the nematic director. The random pinning effect of nematic director and short range disordered smectic
phases were studied extensively for this material by x-ray diffraction techniques \cite{mmt,Germano,Leheny,Sungil,mmt1,mmt2,Freelon}.  
In panel (b) of Fig.\ref{fig1}, the $\rho_s$=0.22 g/cm$^3$ sample
surface is shown. As the aerosil mass density increases the surface profile is found to be rougher than the one in panel (a). This can be seen
from increased color-bar intensity values and the white regions are more pronounced  compared to the white regions of panel (a). The average size of these regions
were measured as $\sim$90 nm. The roughness values are given in Table.\ref{tablo1}. The increase in size of diffused aerosil aggregated regions can be 
seen from  panel (c) to panel (e) where the aerosil mass density increases to $\rho_s$=0.347, 0.49 and 0.647 g/cm$^3$, respectively. These samples , different 
than the ones shown before, are found to be less gel-like, more solid, similar to small pebbles. The biggest change is the surface porosity which became 
so much rougher. Especially for the highest $\rho_s$ 
sample, the aggregation of aerosil form aerosil groups , which is shown by white elongated spheres, $\sim$250 nm in size, meaning 
$\sim$ 35 silica particles. These can be 
seen easily even with the naked eye, shown in Fig.\ref{fig2}. These aerosil groups and their surface appearances were also studied 
in aerosil tablet forms 
\cite{Buscarino, Parvinzadeh}. When we compare these results, the similarities are spectacular and this shows that starting with the $\rho_s$=0.347 sample, 
the volume
and therefore the surface is dominantly composed of aerosils. This also confirms the very short random correlation lengths observed for the high aerosil mass
density samples from x-ray scattering experiments \cite{mmt}. Especially starting with $\rho_s$=0.347 g/cm$^3$ , $\xi_{RD}$ becomes very short and saturates at 
$\sim$ 400, $\sim$ 500 \AA~ for 4 different LCs; 8CB, 8OCB, 10CB, 408 \cite{mmt,mmt1}. This basically is another confirmation using only the surface profile information 
that the random perturbation observed in 
aerosil dispersed LCs are developed through the values of $\rho_s$=0.347 and that the latter is completed at these short length limits. In other words,
if we continue to mix more aerosil in the same volume  of the LC host, the resultant gel will not introduce any other information than these two, three
high density samples. The same conclusion would also be derived from the aerosil group size values given in Table \ref{tablo1}. The typical aerosil
group size becomes almost saturated at $\sim$200 - $\sim$300 nm starting with $\rho_s$=0.347 g/cm$^3$.   
As stated before, Fig.\ref{fig2} shows the real picture from the surface of the $\rho_s$=0.647 sample. The big white colored droplet-like areas are the 
aerosil aggregated regions where the real surface roughness could reach up to almost $\sim$ 10 $\mu$m. This kind of high roughness  could obviously 
only be measured by a profilometer. When we compare this surface with the ones of the $\rho_s$=0.49 g/cm$^3$ sample -which is not shown here- we see that we do not
have any formations of these big white colored clusters. The surface of the $\rho_s$=0.49 g/cm$^3$ sample is much smoother than the $\rho_s$=0.647 g/cm$^3$ sample.
The profilometer and also the AFM measurements showed this numerically as well. 

\begin{figure}
\includegraphics[width=1\linewidth]{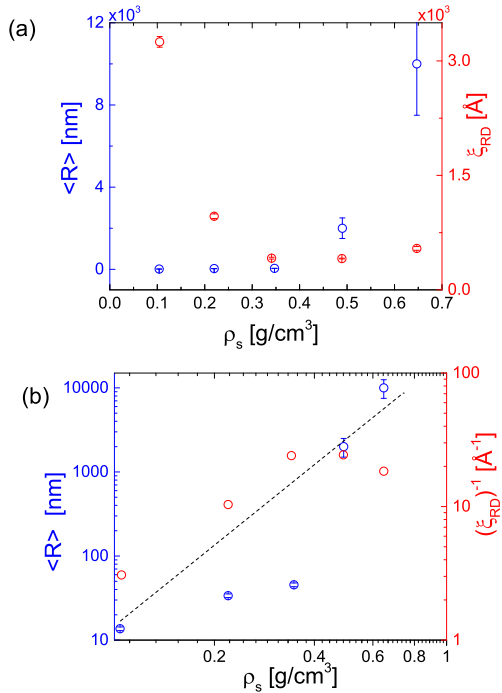}
\caption{\footnotesize Surface roughness and random correlation length versus random disorder strength. Surface roughness; $<R>$ and the correlation
length; $\xi_{RD}$ values
given in Table.\ref{tablo1} are shown with respect to random disorder strength; $\rho_s$. Panel (a) shows this behaviour linearly while panel (b) is drawn to show
it in a log-log plot.
As seen from panel (a), the roughness starts from a low limit and increases logarithmically as the $\rho_s$ values increase. The corresponding behaviour is
the inverse of this for $\xi_{RD}$.
In order to show these two correlated trend together,  panel (b) is created. The inverse random disorder correlation length multiplied with a constant amplitude
is shown in a log-log 
plot with the roughness data. \textbf{The line in this panel is for eye, only}. }
\label{fig3}
\end{figure}

The overall roughness analysis of the whole sample spectrum is shown in Fig.\ref{fig3}. In this figure, the corresponding roughness values are given along 
with the random correlation lengths obtained from previous x-ray scattering experiments \cite{mmt}. In doing so, we aim to show the correlation between the
roughness and  the short ranged ordering in the sample gels on the same graph. In this figure, as the aerosil mass density increases,
so does the roughness $<$R$>$ values while the random correlation length $\xi_{RD}$ decreases. Both behaviour of $<$R$>$ and $\xi_{RD}$ are related to $\rho_s$, exponentially. This is shown in a linear figure in panel (a).
In order to show the same behaviour in a logarithmic figure, we use the inverse $\xi_{RD}$ multiplied with a constant amplitude of 10000. This way,
we show the correlation between $<$R$>$ and $(\xi_{RD})^{-1}$ clearly in logarithmic figure in panel (b). Panel (b) is, in fact, the most important outcome
of this surface study. The increase in aerosil mass density, aka the random field effect, which increases the random pinning of LC nematic orientation 
-so too does the smectic layers- changes the correlation length of smectic-A x-ray diffraction peaks while the same physical information can also be obtained 
from the surface topography of the LC+aerosil gels. In other words, as the $\rho_s$ increases, the sample becomes less correlated and rougher on 
the surface.

\begin{table}
\caption {The length scales used in this study given as a function of random disorder strength; $\rho_s$. The units for $\rho_s$, roughness; $<R>$ , random correlation
lengths; $\xi_{RD}$ and average aerosil group size ; GS are g/cm$^3$, nm , \AA  ~and nm respectively. The $\xi_{RD}$ values were obtained from previous 
8CB+Aerosil and 8OCB+Aerosil x-ray diffraction experiments
\cite{mmt,Sungil,Germano,Leheny}. The $\xi_{RD}$ given in the last column belong to 8CB+Aerosil data except the last two which correspond to $\rho_s$=0.49 and 
$rho_s$=0.647 g/cm$^3$ samples. These are from 80CB+Aerosil and 10CB+Aerosil experiments, respectively \cite{Paul,mmt1,mmt}. The average aerosil group size is measured 
on the pictures manually using the AFM analysis software \cite{AFM}.
The roughness values given for these two samples contain both AFM and profilometer results. }

\label{tablo1}  
\begin{tabular}{| c | c | c | c |c|}
\hline

~$\rho_s$~       &~$wt$\%~       & ~$<R>$~               &~$\xi_{RD}$~    & ~GS~           \\ \hline\hline
~0.104~          &~9.5~          & ~13.7 $\pm$ 0.7~      & ~3250 $\pm$ 70 & ~$\sim$50~           \\ 
~0.22~           &~18~           &~33.7 $\pm$1.7~        & ~965 $\pm$30   & ~$\sim$90~          \\
~0.347~          &~26~           &~45.5 $\pm$2.3 ~       & ~415 $\pm$12   & ~$\sim$140~          \\
~0.49~           &~33~           & ~2000 $\pm\sim$500~   & ~410 $\pm$5    & ~$\sim$210~          \\
~ ~              &~ ~            & ~160 $\pm$8~          & ~  ~           &~ ~                   \\
~0.647~          & ~ 39.5~       & ~10000$\pm\sim$2500~  &~ 543 $\pm$24~  &~$\sim$250~           \\ 
~ ~              &~ ~            & ~180 $\pm\sim$10 ~    &~ ~             &                       \\ \hline                                                                                             

\end{tabular}\\ 
\end{table}

\section{Conclusion and Discussion}
In this study, we investigated the LC+aerosil sample gels using their surface topography. The aerosil dispersion in the LC host creates fractal like regions 
after a diffusion-limited aggregation process. This is seen from the surface topography from the low to high aerosil mass density samples. 
Depending on the aerosil amount in the LC host, we show that there is a net increase in the roughness value which can be used to characterize the
sample surfaces. This result is correlated with the results of previous x-ray scattering experiments. Especially for the random field correlation length 
used to determine the short ranged ordering in LC+aerosil gels, is found to be correlated with the roughness values. The main overall outcome of 
this first study is being able to show that the random field effects on aerosil dispersed LC gels can be modeled using surface roughness. 
We believe this is the first surface study of LC+aerosil gels confirming the existence of quenched randomness as a conclusion.

$Acknowledgement$ 
MR would like to thank Prof. Birgeneau for his support and constructive comments. This study  
is supported by TUBITAK, 115F315.


\begin{thebibliography}{20}
\bibitem{AFM} A scanning prome microscopy made by Nanomagnetics was used with tapping and non-contact mode. The AFM probe tapping frequencies were $\sim$
161 kHz for these scans.
\end{thebibliography}
\end{document}